\documentclass[twoside,12pt]{article}

\usepackage[english]{babel}
\usepackage{alltt}
\usepackage{moreverb}
\usepackage{graphicx}

\newcommand{\foldl}{\emph{foldl}}
\newcommand{\collect}{\emph{collect}}
\newcommand{\Collect}{\emph{Collect}}

\newcommand{\dcl}{\usefont{T1}{cmtt}{m}{n}}
\newcommand{\myemph}[1]{{\dcl #1}}

\newcommand{\programlabel}{}
\newcommand{\programtitle}{}

\newenvironment{programA}[2]
{% ---------- début de l'environnement
\begin{figure}[htbp] 
\renewcommand{\programlabel}{#1}
\renewcommand{\programtitle}{#2}
\hrule
\begin{footnotesize}
\begin{alltt}
    
}{% ---------- fin de l'environnement
\end{alltt}
\end{footnotesize}
\vspace{-0.5cm}
\label{\programlabel}
\caption{\programtitle}
\vspace{0.3cm}
\hrule
\end{figure}
}% End programA

\begin{document}
\pagestyle{myheadings}
\markboth{AADEBUG 2000}{Collecting graphical abstract views}

\title{ Collecting Graphical Abstract Views \\of Mercury Program
Executions
\footnote{In M. Ducass\'e (ed), proceedings of the Fourth
International Workshop on Automated Debugging  
(AADEBUG 2000), August 2000, Munich. 
COmputer Research Repository (http://www.acm.org/corr/), cs.SE/0010038;
whole proceedings: cs.SE/0010035.}}
\author{Erwan~Jahier\\
IRISA,  
Campus Universitaire de Beaulieu,\\
F-35042 RENNES Cedex - France\\
jahier@irisa.fr, www.irisa.fr/lande/jahier}
\date{}
\maketitle

%--------------------------------------------------------------------%
\begin{abstract}
  A program execution monitor is a program that collects and
  abstracts information about program executions. The
  \collect~operator is a high level, general purpose primitive which
  lets users implement their own monitors.  \Collect~is built on top
  of the Mercury trace. In previous work, we have demonstrated how
  this operator can be used to efficiently collect various kinds of
  statistics about Mercury program executions. In this article we
  further demonstrate the expressive power and effectiveness of
  \collect~by providing more monitor examples. In particular, we show
  how to implement monitors that generate graphical abstractions of
  program executions such as proof trees, control flow graphs and
  dynamic call graphs. We show how those abstractions can be easily
  modified and adapted, since those monitors only require several
  dozens of lines of code. Those abstractions are intended to serve
  as front-ends of software visualization tools.  Although
  \collect~is currently implemented on top of the Mercury trace, none
  of its underlying concepts depend of Mercury and it can be
  implemented on top of any tracer for any programming language.

\end{abstract}

%--------------------------------------------------------------------%
\section{Introduction}

A \emph{program execution monitor} is a program that collects and
abstracts information about program executions. The monitoring
functionalities of existing systems are built on top of ad hoc
instrumentations. Most of them are implemented by subtle
modifications of the runtime system; therefore, implementing such
monitors require an in-depth knowledge of the system.  The best
people to implement these instrumentations are generally the
implementors of the compiler.  They, however, cannot decide which
data to gather. Indeed, hundreds of variants can be useful and only
end-users know what they want.

\Collect~is a high level, general purpose, \foldl-like operator which
lets users implement their own monitors. We have demonstrated
in~\cite{jahier99d} how this operator can be used to collect various
kinds of statistics about Mercury program executions such as counting
the number of predicate calls, the number of events for each event
type (port), or the number of events at each depth. We have also
showed how it is possible to perform test coverage ratio; this
information is useful to assess the quality of a test set. The aims
of these examples were twofold: demonstrating the expressive power of
\collect~and the efficiency of the resulting monitors.
%
% Contribution 1: more abstract views --> flexibility and efficiency
In this article, we propose more program abstractions implemented
with \collect.  The goal is to further assess the expressive power of
\collect~in a pragmatic way by implementing a wider range of monitors
and to check that the resulting monitors are still reasonably
efficient.  The monitors described in this article requires slightly
more programming effort than the ones in \cite{jahier99d}.
%
% Contribution 2: a basis as a front end for visualization tools
In particular, we show how to implement monitors that generate
graphical abstractions of program executions such as proof trees,
control flow graphs and dynamic call graphs. We show how those
abstractions can be modified and adapted.  We believe that
\collect~could be the basis of software visualization tools
\cite{stasko98}.
%
% Contribution 3: A tutorial to implement monitors with collect
This article also aims to be a tutorial about how to implement (Mercury)
program monitors with \collect.

All the monitors given in this article are run under the Morphine
trace analysis system. Morphine~\cite{morphine-ref} is a Prolog
interpreter enhanced with primitives (\collect\ included) to
communicate with a Mercury program's execution.  Morphine is a fully
programmable command line interface for interactively monitoring and
debugging Mercury program executions.
The use of \collect~is independent of other Morphine concepts though;
we only use the Prolog part of Morphine to post-process monitor's
results. All the examples are verbosely paraphrased so no knowledge
about Mercury or Prolog should be required to understand them.
Section \ref{mercury-and-collect} gives a quick overview of the
language and the trace system of Mercury, as well as an informal
presentation of \collect.  Sections~\ref{cfg-section},
\ref{dcg-section}, and~\ref{pt-section} show how to implement
monitors that generate control flow graphs, dynamic call graphs and
proof trees respectively. Section~\ref{merge} describes how monitors
can be merged.  Section~\ref{performance} discusses performance issues
and Section~\ref{related} related work.

%--------------------------------------------------------------------%

\section{Flexible and efficient monitoring of Mercury}
\label{mercury-and-collect}

The \collect~operator \cite{jahier99d} is a generic primitive
designed to let users implement easily efficient Mercury program
monitors. In this section, we give a brief review of the Mercury
programming language and of the \collect~monitoring operator.

\subsection{A brief introduction to Mercury}
\label{mercury-pres}

Mercury \cite{somogyi96} is a purely declarative, logical and
functional programming language.  Its syntax is very similar to the
one of Prolog. The main difference with Prolog is that users must
declare the type, the mode and the determinism of predicates (and
functions) they define. These declarations let the Mercury compiler
produce very robust and efficient code.

\begin{programA}{mercury-ex}{The Mercury predicate queen/2}
    :- pred queen(list(int)::in, list(int)::out) is nondet.
    queen(Data, Out) :-
            qperm(Data, Out),
            safe(Out).
\end{programA}

Figure \ref{mercury-ex} shows an example of Mercury code.  It is a
predicate of a Mercury program that solves the well known $n$ queens
problem. This program is given in Appendix 1.  The first line is the
type and mode declaration of predicate \myemph{queen/2}\footnote{
  \myemph{queen/2} denotes a predicate named \myemph{queen} of arity
  \emph{2}.}. This line states that the first argument of
\myemph{queen/2} is a list of integers and this argument is an input
(\myemph{in}). It also states that its second argument is a list of
integers and it is an output (\myemph{out}). The \myemph{nondet}
determinism marker means that this predicate can have any number of
solutions.  Actually, it has two solutions.  The list of integers
codes the board. The predicate \myemph{qperm/2} generates all the
possible configurations by producing all the permutations of the list
of integers. Then \myemph{safe/2} checks that a given configuration
is a solution, namely, that there is no more than one queen per
diagonal.

\paragraph{The Mercury trace}

A \emph{trace} is a sequence of events. An \emph{event} is a tuple of
event attributes.  An \emph{event attribute} is an elementary piece
of information that can be extracted from the current state of
particular points of the program execution.  The program points of
the Mercury trace are chosen according to a trace model that is
called Byrd's box model \cite{byrd80}: a \emph{call} event is
generated when a predicate is called; a \emph{exit} event is
generated if it succeeds; a \emph{fail} event is generated if it
fails; a \emph{redo} event is generated if the execution backtracks
on a predicate to see if it can produce other solutions.  Actually,
the Mercury trace is an extended version of Byrd's box model because
events are also generated when the execution enters a branch of a
disjunction or of an if-then-else. In the following, theses events
are called \emph{internal} and the Byrd's event are called
\emph{external}.  The complete list of Mercury event attributes is
given in Appendix~2.

%--------------------------------------------------------------------%
%\section{A monitoring with a trace processing operator}
\subsection{The \collect~monitoring operator}
\label{collect}

Debugging and monitoring can be seen as a list of events processing
activity.  The standard functional programming operator \emph{foldl}
encapsulates a simple pattern of recursion for processing
lists\footnote{The \emph{foldl} operator takes as argument a
function, a list, and an initial value of an accumulator; it outputs
the final value of the accumulator; this final value is obtained by
applying to the function the current value of the accumulator and
each element of the list successively; the \emph{l} at the end of
\emph{foldl} means that this list is processed from left to rigth.}.
As demonstrated by Hutton \cite{hutton99}, \emph{foldl} has a great
expressive power for processing lists.  Therefore \emph{foldl} is
likely to be a good abstraction to implement dynamic analysis tools.

However, implementing monitors by collecting the whole execution
trace into a list of events, and then applying a \foldl\ to that list
would be far too inefficient. It would require to create and process
a list of possibly millions of events. To implement efficient
monitors, runtime information needs to be collected and analyzed on
the fly. The primitive \collect~\cite{jahier99d} is a \emph{foldl}
operator which is directly plugged into the trace system.
First, a global variable is created and initialized.  Then, whenever
an event occurs, the \collect~interface is called instead of the
standard debugger. The \collect~interface calls the filtering
predicate which updates the global variable and then gives control
back to the execution.

It is important to note here that for performance reasons, there is
no coroutining between different Operating System (OS) processes (the
program and its monitor) but only procedure calls that update a
global variable. This design decision was made to avoid those
expensive OS level context switches induced by coroutining that the
\collect~operator was initially designed (see related work section).

For the time being, the only implementation of \collect~we have is
done on top of the Mercury trace. The trace component of
the Mercury system has been extended so that it is able to call the
\collect\ interface rather than the Mercury debuggers~\cite{somogyi99}. 
To implement monitors using \collect, users need to give an initial
value to the accumulator by defining a Mercury predicate named
\myemph{initialize/1}, and to update the accumulator at each event by
defining a Mercury predicate named \myemph{filter/4}. Since Mercury
is a typed language, users also need to define the type of the
collecting variable \myemph{collected\_type}.

\begin{programA}{use-collect}{What the user needs to type to define a monitor 
with \collect} 
    % 1 - Importation of Mercury library modules:
    :- import_module <{\sl Mercury modules} >.

    % 2 - Definition the type of the collecting variable:
    :- type collected_type == <{\sl A Mercury type} >.

    % 3 - Initialization of the collecting variable:
    initialize(Accumulator) :-
            <{\sl Mercury goals which initialize the collecting variable} >.

    % 4 - Updating of the global variable:
    filter(Event, AccumulatorIn, AccumulatorOut, StopFlag) :-
           {<\sl Mercury goals which update the collecting variable} >.
\end{programA}

\noindent
This is summed up in Figure~\ref{use-collect}; predicates
\myemph{initialize/1} and \myemph{filter/4} should follow the following
Mercury declarations:
\begin{alltt}
  :- mode initialize(collected\_type::out) is det.
  :- pred filter(event::in, collected\_type::in, 
      collected\_type::out, stop\_or\_continue::in) is det.
\end{alltt} 
The type \myemph{event} is a structure which contains all the Mercury
event attributes.  To access those attributes, the monitor designer
can use attribute accessor functions whose prototypes are of the
form:
\begin{verbatim}
     :- func <attribute_name>(event::in) = <attribute_type>::out.
\end{verbatim}
\noindent
For example, the function \myemph{depth(Event}) returns the depth of
the event \myemph{Event} (the full list of \myemph{attribute\_name}
is given in Appendix 2). The fourth argument of \myemph{filter/4} is
a binary flag that can be set to \verb+stop+ or \verb+continue+
depending on whether or not one wants to stop the monitoring process
before the end of the execution is reached.
The current front-end of \collect~is a Prolog interpreter. Having a
full programming environment is very useful to post-process the
results of the monitor.  If a file called \myemph{my\_monitor}
contains a definition of \myemph{initialize/1} and \myemph{filter/4},
then the call \myemph{collect(queens, my\_monitor, Result)} binds
\myemph{Result} with the result of the monitor specified in
\myemph{my\_monitor} to the Mercury program \myemph{queens}. For
example, Figure~\ref{count-call-ex} contains the full code of a
monitor that counts the number of predicate calls.  If this code is
in a file called \verb+count_call+, then the query
\myemph{collect(queens, count\_call, Result)} will bind \verb+Result+
to the number of predicate calls that occurs during the execution of
the program \verb+queens+.

\begin{programA}{count-call-ex}{count\_call: a monitor that counts the 
    number of predicate calls}
    :- import_module int.
    :- type collected_type == int.

    initialize(0).
    filter(Event, Cpt0, Cpt, continue) :-
            ( if port(Event) = call then
                    Cpt = Cpt0 + 1
            else
                    Cpt = Cpt0 ).
\end{programA}

Here is a line by line description of the code of
Figure~\ref{count-call-ex}. To do that, here and in the following
descriptions of monitors defined in this article, we successively
describe each of the four points that need to be addressed to define
a monitor using \collect. (1) Importing necessary Mercury modules:
here, we only need to import the library module \myemph{int} that
defines everything that is concerned with integers.  (2) Defining the
type of the collecting variable: here, it is an integer. (3)
Initializing the collecting variable: here, it is initialized to~$0$.
(4) Defining the filtering predicate: here, it increments the global
variable whenever the current event port attribute is \myemph{call}.

\begin{figure}[htbp] 

\hrule \vspace{0.3cm}
\includegraphics[width=\textwidth,height=2.5cm]{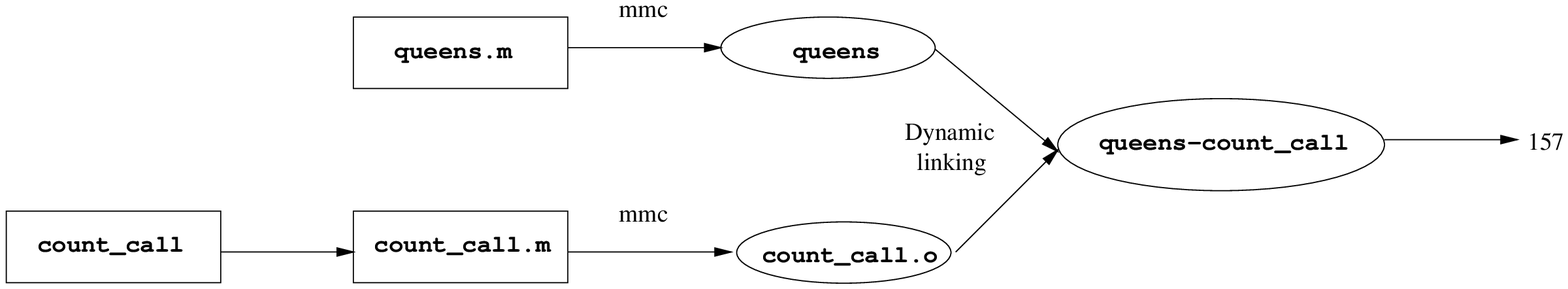}
\caption{the various involved components and
their relation when the user invoke the command
\myemph{"collect(queens, count\_call, Result)"}.}
\label{components}
\vspace{0.3cm}
\hrule
\end{figure}

Figure~\ref{components} shows the various involved components and
their relation when the user invoke the command
\myemph{``collect(queens, count\_call, Result).''} at the Morphine
prompt. Square boxes represent source files. Circle boxes represent
object and executable files.  The file containing the definition of
the monitor, \myemph{count\_call}, is transformed by Morphine into a
Mercury module.  The arrows labeled with \verb+mmc+ denote a call to
the Melbourne Mercury compiler.  The compilation of
\verb+count_call.m+ and \verb+queens.m+ is only done if necessary,
i.e., if something has changed in the source code since the last time
it was compiled. The executable file \myemph{queens-count\_call} 
is obtained by dynamically linking  the executable file
\myemph{queens} and the objet file \myemph{count\_call.o}. The output
of this program, 157, is unified with the logical variable
\myemph{Result} at the Morphine prompt.

%--------------------------------------------------------------------%
\section{\emph{Collect}ing graphical abstract views}
\label{graphs}

In this section, we show how to generate several kinds of program
execution abstract views based on graphs. All the graphs of this
section are visualized with the dot drawing tool~\cite{koutsofios91}.
The final objective is to have more sophisticated back-ends such as
the visualization tools presented in~\cite{stasko98}. The point of
this section is not to provide an exhaustive set of graphical
abstractions. The point is to show how easy it can be to implement
them and, more interestingly, how easy it is to get a variant of an
existing abstraction. Indeed, different visualization tools often
need slightly different images of the execution.
The full code that implements the production of these graphs is
available in the control\_flow scenario of the Morphine
distribution\footnote{http://www.irisa.fr/lande/jahier/download.html}.

%--------------------------------------------------------------------%
\subsection{Control flow graphs}
\label{cfg-section}

\begin{figure}[htbp] 
\hrule \vspace{0.3cm}
\vspace{0.3cm}
\hspace{3cm}
\includegraphics[width=6 cm,height=6cm]{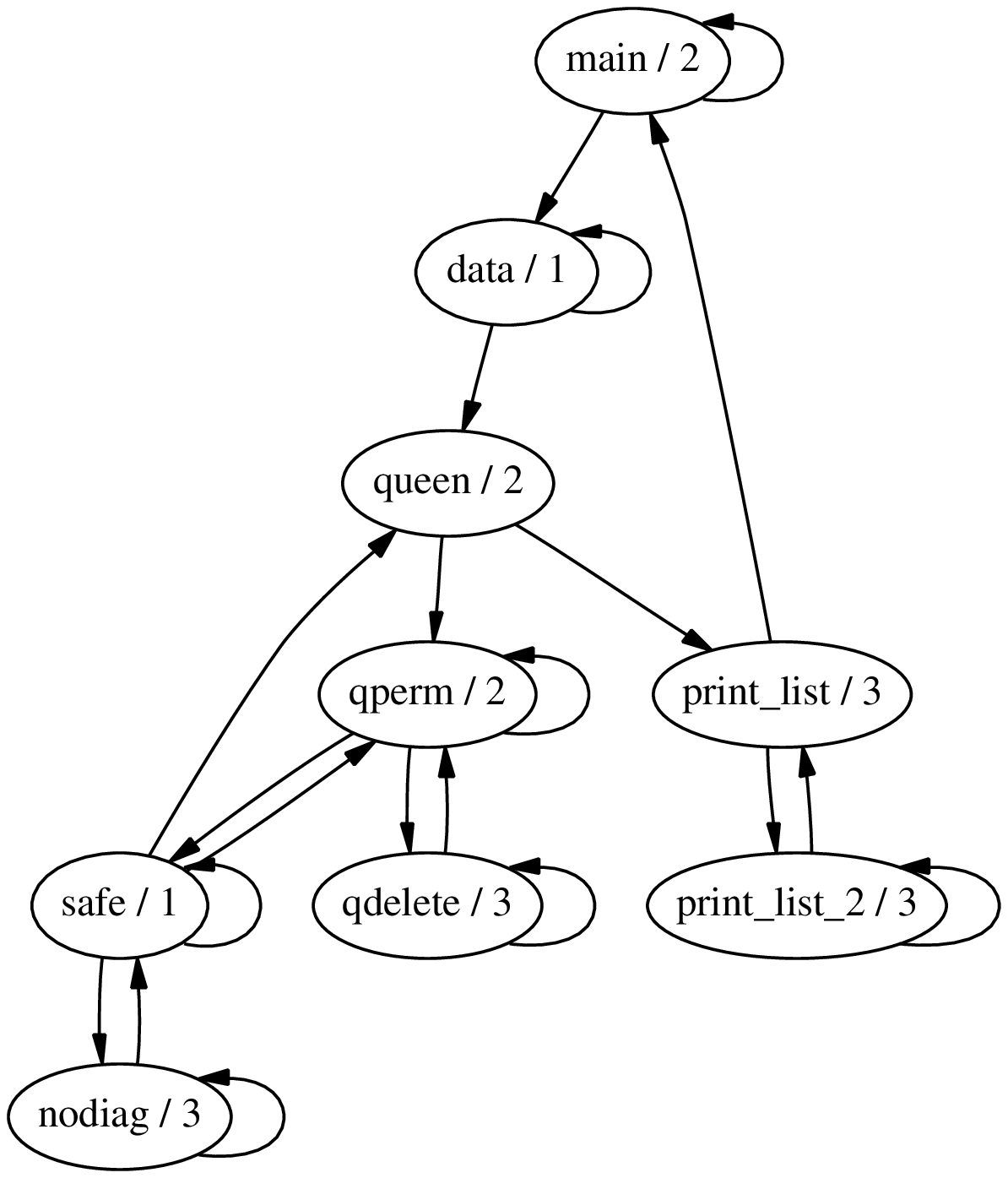}
\caption{The control flow graph of $n$ queens program}
\label{cfg-eps}
\vspace{0.3cm}
\hrule
\end{figure}

We define the \emph{control flow graph} of a logic program execution
as the directed graph where: nodes are predicates of the program; and
arcs indicate that the program control passed from the origin to the
destination node during the execution.  Control flow graphs are
useful execution abstractions for users to understand what a program
actually do.  They are also the basis of a lot of dynamic analyses.
The control flow graph of the $n$ queens program is given in
Figure~\ref{cfg-eps}. We can see that, during the program execution,
the control passed from predicate \myemph{main/2} to predicate
\myemph{data/1}, from predicate \myemph{data/1} to predicate
\myemph{data/1} (recursive call) and predicate \myemph{queen/2}, etc.
An implementation of a monitor that performs such a graph is given in
Figure~\ref{cfg-source}.

\begin{figure}[htbp]
\begin{small}
\hrule
\vspace{0.3cm}
\begin{listing}[1]{1}
:- import_module set.

:- type predicate ---> proc_name/arity.
:- type arc ---> arc(predicate, predicate).
:- type graph == set(arc).
:- type collected_type ---> collected_type(predicate, graph).

initialize(collected_type("user"/2, set__init)).

filter(Event, Acc0, Acc, continue) :-
    Port = port(Event),
    ( if (Port = call ; Port = exit ; Port = fail ; Port = redo) then
         Acc0 = collected_type(PreviousPred, Graph0),
         CurrentPred = proc_name(Event) / proc_arity(Event),
         Arc = arc(PreviousPred, CurrentPred),
         set__insert(Graph0, Arc, Graph),
         Acc = collected_type(CurrentPred, Graph)
    else    
         % internal events
         Acc = Acc0 
).
\end{listing}
\hrule

\caption[Moniteur qui calcule des graphes de flot de
contrôle]{Monitor that calculates control flow graphs
\label{cfg-source}}
\end{small}
\end{figure}

\noindent
Monitor of Figure~\ref{cfg-source} is defined as follows.  (1) The
\verb+set+ module of the Mercury library is imported.  (2) Graphs are
encoded by a set of arcs, and arcs are terms made up with two
predicates.  The collecting variable is made of a predicate and a
graph. The predicate is used to remember the previously visited node.
(3) The collecting variable is initialized with the predicate
\verb+main/2+, the top level predicate of every Mercury program,
and the empty graph (\verb+set__init/0+).  (4) For every external
event, we insert in the graph (\verb+set__insert/3+) an arc from the
previous predicate to the current one.
When the program execution has terminated, we post-process the result
with dot \cite{koutsofios91}, a system that takes a graph description
and that displays a pretty post-scripted version of it.
Figure~\ref{cfg-eps} is the output of such a post-processed result
with the monitor of Figure~\ref{cfg-source}.  This post-processing
stage only requires a few dozen lines of Prolog code.
In our definition of control flow graph, the number of times each arc
is traversed is not given. Even if the control passed between two
nodes more than once, only one arc is represented.  One can imagine
variants where, for example, arcs are labeled by a chronological
counter.
%, or by the number of times each arc is exercised.  
The corresponding monitor can be implemented by replacing
\myemph{arc(predicate, predicate)} by \myemph{arc(predicate, chrono,
  predicate)} in the type definition of \myemph{arc}, and replacing
the goal \myemph{Arc = arc(PreviousPred, CurrentPred)} by \myemph{Arc
  = arc(PreviousPred, chrono(Event), CurrentPred)} in the definition
of \myemph{filter/4}.

%--------------------------------------------------------------------%
\subsection{Dynamic call graphs}
\label{dcg-section}

\begin{figure}[htbp]
\hrule
\vspace{0.5cm}
\hspace{3cm}
\includegraphics[width=8 cm,height=5cm]{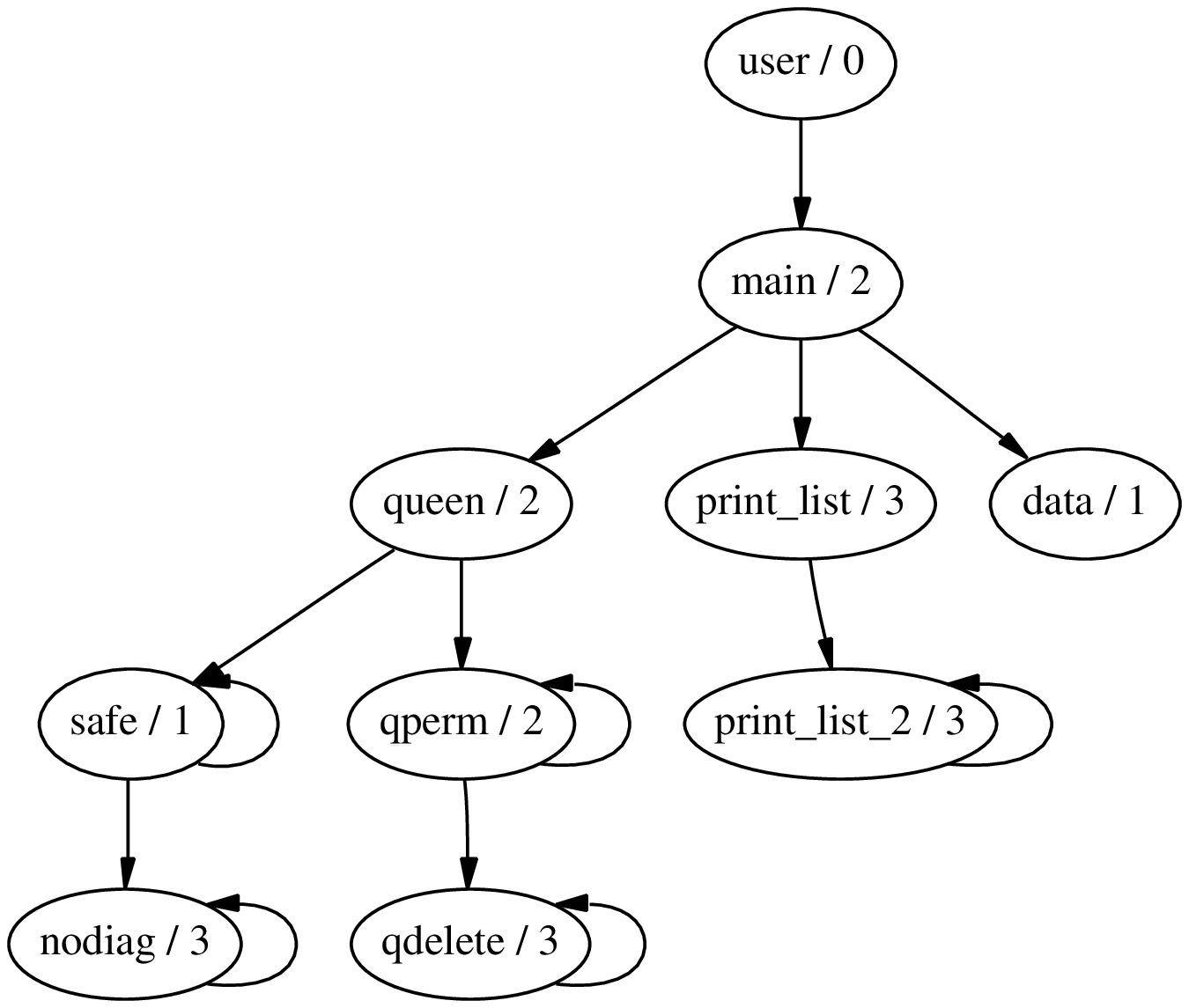}
\caption{The dynamic call graph of $n$ queens program}
\label{dcg-eps}
\vspace{0.3cm}
\hrule
\end{figure}

We define the \emph{dynamic call graph} of a logic program execution
as the sub-graph of the (static) call graph composed of the nodes
that has been exercised during the execution.  In other words, it is
an execution slice of the program call graph.  For example, we can
see that predicate \myemph{main/2} called predicates \myemph{data/1},
\myemph{queen/2} and \myemph{print\_list/2}.
An implementation of a monitor that performs this graph
is given in Figure~\ref{dcg-source}.

\begin{figure}[htbp]
\begin{small}
\hrule
\vspace{0.3cm}
\begin{listing}[1]{1}
:- import_module set, stack.

:- type predicate ---> p(proc_name, arity).
:- type arc ---> arc(predicate, predicate).
:- type graph == set(arc).
:- type collected_type ---> ct(stack(predicate), graph).

initialize(ct(Stack, set__init)) :-
        stack__push(stack__init, p("user", 0), Stack).

filter(Event, ct(Stack0, Graph0), Acc, continue) :-
        Port = port(Event),
        CurrentPred = p(proc_name(Event), proc_arity(Event)),
        update_call_stack(Port, CurrentPred, Stack0, Stack),
        ( ( Port = call ) ->
                PreviousPred = stack__top_det(Stack0),
                set__insert(Graph0, arc(PreviousPred, CurrentPred), Graph),
                Acc = ct(Stack, Graph)
        ; 
                Acc = ct(Stack, Graph0) ).

:- pred update_call_stack(trace_port_type::in, predicate::in, 
        stack(predicate)::in, stack(predicate)::out) is det.
update_call_stack(Port, CurrentPred, Stack0, Stack) :-
        ( ( Port = call ; Port = redo ) ->
                stack__push(Stack0, CurrentPred, Stack)
        ; ( Port = fail ; Port = exit ) ->
                stack__pop_det(Stack0, _, Stack)
        ; % internal ports      
                Stack = Stack0 ).
\end{listing}
\hrule

\caption[Moniteur qui calcule des graphes d'appels dynamiques]
{Monitor that calculates dynamic call graphs}
\label{dcg-source}
\end{small}
\end{figure}

Here is a line by line description of the code of
Figure~\ref{dcg-source}.  (1) Library modules \myemph{set} and
\myemph{stack} are imported.  (2) In order to define this monitor, we
use the same data structures as for the previous monitor, except that
we replace last exercised predicate by the whole call stack in the
collected variable type.  This call stack is computed on the fly.
(3) The stack and the set of arcs are initialized to the empty stack
and to the empty set respectively.  (4) At call events, we insert an
arc from the previous predicate to the current one and we push
(function \verb+stack__push/2+) the current predicate on the stack.
At redo events, we only update the call stack by pushing on it the
current predicate.  At fail and exit events, we remove the top
element of the stack (predicate \verb+stack__pop_det/3+).
The post-processed result of the execution of this monitor on the $n$
queens program is given in Figure~\ref{dcg-eps}.

In the current implementation of \collect, the call stack is not
passed as an event attribute. The reason for that is that the call
stack can be very large, which would slow down the \collect~monitors.
Another reason is that, as demonstrated in this example, it is very
easy to reconstruct this information.  It is also interesting to note
that the impact on the performance of handling the stack on the fly
as we do here is hardly noticeable.

%--------------------------------------------------------------------%
\subsection{Proof Trees}
\label{pt-section}

Another widely used program abstract view in the logic programming
community are proof trees. A \emph{proof tree} of a program execution
is the dynamic call graph where all the fail nodes are omitted.
Thus, for example, a failing request produces an empty proof tree.
We do not give the code of the monitor that implements the proof
tree, but rather briefly explains how it can be done with \collect.
The idea is to maintain a table of proof trees and a table of goal
immediate successors both indexed by goal numbers (each goal is
uniquely defined by its goal number). When predicates successfully
exit, we construct the proof tree of the current goal.  We can do
that because at that port, we have the whole list of the current goal
immediate successors and we know the proof trees of each of these
successors. At redo ports, those proof trees are removed from the
table of proof trees. In order to calculate the list of immediate
successors, we also need to maintain a stack of goal numbers, in
exactly the same manner as in the two previous monitors.

\begin{figure}[htbp] 
\hrule
\vspace{0.3cm}
\includegraphics[width=\textwidth,height=8cm]{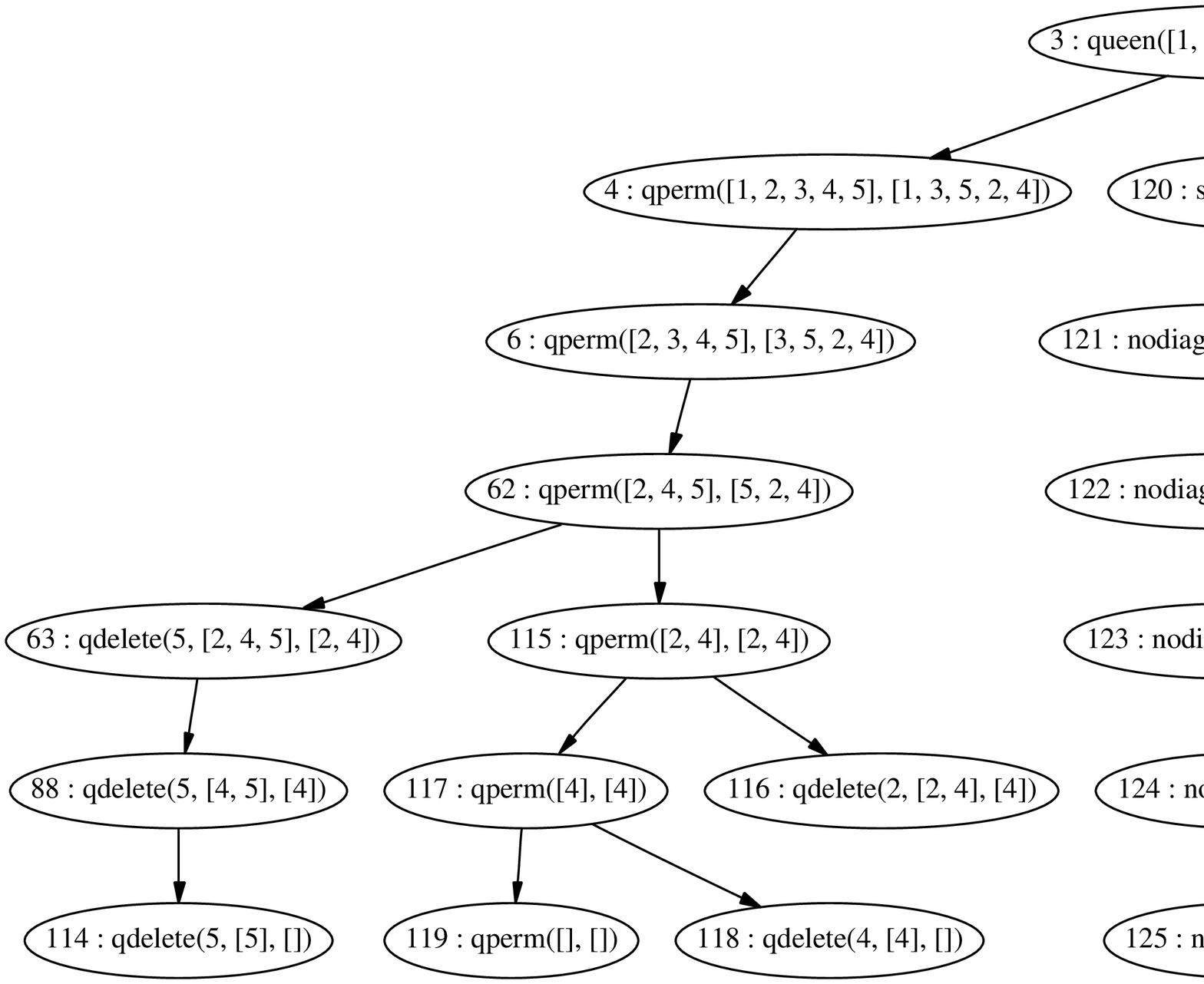}
\caption{The proof tree of $n$ queens program}
\label{pt-eps}
\vspace{0.3cm}
\hrule

\end{figure}

The post-processed result of the execution of this monitor on the
$n$~queens program is given in Figure~\ref{pt-eps}.  It is also
possible to construct SLD-trees with the same kinds of monitors, but
we lack the necessary space to describe it here\footnote{the source
  code of all those monitors, included SLD-tree, is available at the
  Morphine web site http://www.irisa.fr/lande/jahier/download.html}.
We have not included the predicate arguments in the graphs node, but
that could have easily been done.  We could add all the event
attributes in the graph nodes actually. But then we would really need
a visualization tool back-end that, for example, would display all
those informations on request by clicking on nodes.

%--------------------------------------------------------------------%
\section{Merging monitors}
\label{merge}
\label{compose}

\begin{programA}{set-monitor}{A set of monitor indexed by $i \ in 
\{1, ...,n \}$}
    :- type collected_type ---> type_i.

    initialize(Ci) :- initialize_i(Ci).
    filter(Event, CiIn, CiOut, continue) :- filter_i(Event, CiIn, CiOut).
\end{programA}

A nice property of \collect~is that it implements monitors that can
easily be merged. All the monitors can therefore collect their data
in only one program execution.  Indeed, consider the $n$ monitors of
Figure~\ref{set-monitor} where: $i$ is in $\{1, ..., n\}$;
\verb+type_i+ is an arbitrary Mercury type; \verb+initialize_i+ a
predicate that outputs a term of type \verb+type_i+; and
\verb+filter_i+ a predicate that takes an event, a term of type
\verb+type_i+ and outputs a term of type \verb+type_i+.  We suppose
that there are no name clashes between \verb+Ci+, \verb+CiIn+ and
\verb+CiOut+. Then all those monitors can be merged as shown in
Figure~\ref{monitor-merged}.

\begin{programA}{monitor-merged}{The monitor obtained by merging the $n$
monitors of Figure~\ref{set-monitor}}
    :- type collected_type ---> union(type_1, ..., type_n).

    initialize(CollectVar) :-
            CollectVar = union(C1, ..., Cn),
            initialize_1(C1),
            ...
            initialize_n(Cn).

    filter(Event, CollectVarIn, CollectVarOut, continue) :-
            CollectVarIn = union(C1In, ..., CnIn),
            filter_1(Event, C1In, C1Out),
            ...
            filter_n(Event, CnIn, CnOut),
            CollectVarOut = union(C1Out, ..., CnOut).
\end{programA}

The collecting variable type of the resulting monitor is a functor
with arguments the $n$ monitors collecting variable types. The
initialization (resp.  filtering) predicate successively initializes
(resp. updates) each collecting variable using the initialization
(resp. filtering) predicate of the sub-monitors. This can easily be
done automatically.

%--------------------------------------------------------------------%
\section{Performance issues}
\label{performance}

We have seen in the previous sections how it is possible for users,
without any knowledge about the Mercury trace system, to implement
their own monitors.  This genericity has as price: efficiency.  In
this section, we try to exhaustively examine the source of
performance overheads of our approach compared with hand-crafted
ad-hoc monitors implemented inside the compiler.

\paragraph{Granularity of the instrumentation} 
The principal source of overhead is due to the fact that not all the
monitors need such a fine grained instrumentation as the trace system
have.  The only control we have over the granularity of the
instrumentation is the possibility to generate only external events
when compiling programs. In order to assess this overhead, we have
compared the execution times of Mercury programs executed normally
with programs executed within the control of the trace system.  We
mean by executed under the control of the trace system that, at each
event, the trace system is called and then simply returns. We have
measured a slowdown of around a factor of two if internal events are
not generated, and a factor of four if they are generated.

\paragraph{Unused event attributes} 
A second source of overhead is due to the fact that we systematically
pass to \myemph{filter} all event attributes, even if they are not
used. Actually, this is not really a problem since it is possible to
dynamically choose the event attributes that are available in the
event structure. In the current implementation of \collect, it is
already the case for the list of live arguments.
Indeed, since this attribute can be very large, we want to avoid the
cost of handling it when it is not necessary.  To assess this source
of overhead for the other attributes, we have measured an
implementation of \collect~that handles all the event attributes
versus a version that handles none of them (leading to monitors that
can not do anything useful but counting the number of events).  The
slowdown was smaller than 10\%.

\paragraph{Scaling up the approach}
Performance problems might occur when the monitoring data become very
large. A possible solution is to bufferize the data collecting by
sending intermediate data to the Morphine process every N events.
This is possible thanks to the collect fourth argument flag, which
allows to stop the monitoring process at any events. Then Morphine,
which is based on a full existing Prolog programming environment, can
manage the analysis of the collected data and then start a new
\collect\ call to finish the execution monitoring. Moreover, on a two
processors machine, we can even process those partially collected
data asynchronously and thus perform the collecting and the analysis
steps in parallel.

%--------------------------------------------------------------------%

\section{Related Work}
\label{related}

\paragraph{Programmable debuggers}
Ducass\'e designed Coca \cite{ducasse99c} and Opium \cite{ducasse99},
programmable debuggers for C and Prolog respectively.  Coca and Opium
are based on a Prolog interpreter plus an handful of coroutining
primitives connected to the trace system. Those primitives allow the
Prolog interpreter to communicate with the debugged program. Coca and
Opium are full debugging programming languages in which all classical
debuggers commands can be implemented straightforwardly. However, it
appears that the set of coroutining primitives of Coca and Opium are
not well suited for monitoring.  All the monitors implemented with
\collect\ can easily be implemented with this set of primitives, but
the resulting monitors require too much Operating System level
context switches and too much socket traffic between the program and
the monitor.  With program of several million of execution events,
such monitors are several orders of magnitude slower than their
counterparts that use \collect~\cite{jahier99b}.

The \collect\ primitive does not only extend the Coca/Opium
primitives in terms of efficiency, but also in terms of
expressiveness. Using the existing primitives to implement monitors,
one typically duplicate the code that (1) makes the execution move
one event forward, (2) checks if the end of the execution is reached.
Those two steps are automatically done when using \collect.  In other
words, the expressiveness improvement between \collect\ and the
existing primitives of Coca/Opium is the same as the improvement we
have between using \foldl\ and processing lists manually.
As a matter of fact, \collect\ can be seen as a generalization of
Opium/Coca coroutining primitives since they can all be implemented
with it.

\paragraph{Automated development of monitors}
Jeffery and al. designed Alamo~\cite{jeffery98}, an architecture that
aims at easing the development of monitors for C programs.  As in our
approach, their monitoring architecture is based on events filtering
and monitors can be programmed. Their system deals with trace
extraction whereas we rely on an already available tracer; this saves
us from a low level and tedious task which has already been done and
optimized. On the other hand, we do not have the control on the
information available in the trace. Note however that lacking
information can sometimes be recovered as we did for example to
perform the call stack.  Moreover, to avoid code explosion, they
perform part of the events filtering at compilation time. Hence they
need to recompile the program each time they want to execute another
monitor whereas we only need to dynamically link the monitor to the
monitored program. Alamo and the monitored program are running in
coroutining, but within the same address space. Alamo has therefore
less problems of efficiency than Coca and Opium for monitoring.
Eustace and Srivastava developed Atom~\cite{eustace95},
a system that also aims at easing the building of monitors. The
difference with Alamo is that monitors are implemented with procedure
calls and global variables which is much more efficient than
coroutining.  However, the language Atom provide is far less
expressive than the one of Alamo.  Alamo and Atom have influenced the
design of \myemph{collect} and we tried to take the best of both; a
full and high level programming language implemented by procedure
calls.

Kishon and al.~\cite{kishon95} use a denotational and operational
continuation semantics to formally define monitors for a simple
functional programming language. The kind of monitors they define
are profilers, debuggers, and statistic collectors. From the
operational semantics, a formal description of the monitor, and a
program, they derive an instrumented executable file that performs
the specified monitoring activity. The semantics of the original
programs is preserved. Then, they use partial evaluation to make
their monitors reasonably efficient.
The main disadvantage with this approach is that they are rebuilding
a whole execution system from scratch, without taking advantage of
the existing compiler.  We strongly believe that it is important to
have the same execution system for debugging and for producing the
final executable program. As noted by Brooks and al.~\cite{brooks92},
some errors only occur in presence of optimisations, and vice versa;
some programs can only be executed in their optimized form because of
time and memory constraints; when searching for ``hot spots'', it is
better to do it with the optimized program as lots of things can be
optimized away; and finally, sometimes, the error comes from the
optimisations themselves.

\paragraph{Efficient monitoring}
Patil and Fisher \cite{patil97} tackles the problem of performance
monitoring by delegating the monitoring activities to a second
processor that they call a shadow processor. Their approach is very
efficient; the monitored is nearly not slowed down. However, the set
of monitoring commands they propose cannot be extended.

\paragraph{Invariant Detection}
Explicitly stated program invariants can help users to identify
program properties which must be preserved when modifying code.
Invariant discovery is generally done statically \cite{cousot77}.
However, static analyses miss true but uncomputable properties and
properties that depend of the program inputs.  Ernst and al.
investigate a complementary approach~\cite{ernst99} that consists of
dynamically detecting program invariants.  The idea is to run
instrumented versions of programs on a sufficiently large set of test
cases, and then examine the values they compute, looking for patterns
and relationships among those values.  Useless invariants are
filtered~\cite{ernst00}. A prototype implementation, Daikon,
demonstrates the feasibility of this approach. Despite its intrinsic
unsoundness, Ernst and al.  report that dynamic invariant discovery
can be very useful in practice.  We believe that it could be an
interesting application of \collect.

%--------------------------------------------------------------------%
\section{Conclusion}

For a given monitor, provided that the whole necessary information is
in the trace, (1) is it always \emph{possible} to implement a given
monitor?  (2) is it always \emph{easy} to implement it?  (3) is it
always possible to implement it \emph{efficiently}?
Since it is possible to collect the whole execution trace, the answer
to the first point is yes. This would be the most inefficient way of
implementing monitors though as the trace can be huge.  
The second point is more difficult to assess. However, we believe
that \collect\ has the rigth level of abstraction to allow that.
Processing a trace with \collect\ is the same as processing a list
with a \foldl\ operator; and the \foldl\ operator is very expressive,
as argued by Hutton~\cite{hutton99}. Indeed, processing lists using a
\foldl\ operator has significant advantages over processing lists
manually.
This article contributes to give an experimental assessment of the
second point by demonstrating how easy a wide range of monitors can
easily be implemented with \collect.
With regards to the third point, the measurements we made let us
believe that the cost of monitors implemented with \collect\ is
acceptable. We believe that all the monitors implemented with
\collect\ executes in the same order of magnitude of time as their
hand-crafted counterparts.

The choice of Mercury to implement \myemph{initialize} and
\myemph{filter} is arbitrary. The reasons to use Mercury in this
context is that people who want to monitor Mercury programs are very
likely to be Mercury users. Moreover, since \myemph{filter} will be
called possibly millions of times, it makes sense to use a highly
optimized compiler such as the one of Mercury.

%--------------------------------------------------------------------%
\section*{Acknowledgments}
 
%\begin{small}
I warmly thank Mireille Ducass\'e and the anonymous referees for
their fruitful comments, Fergus Henderson for his suggestion to add a
flag as fourth argument of filter to be able to stop the monitoring
process before the end of the execution, and the members of the
Mercury project of the university of Melbourne for the low level
tracer and their efficient help.
%\end{small}

%--------------------------------------------------------------------% 
\begin{small}
%\bibliographystyle{plain} 
%\bibliography{/udd/jahier/rubricabrac/tex/mycommontex}

\end{small}

%--------------------------------------------------------------------% 
\appendix
\section*{Appendix 1 - The n-queens program in Mercury}
\vspace{0.3cm}
\begin{minipage}[t]{.55\linewidth}
\begin{footnotesize}
\textwidth 16cm

\begin{alltt}
:- module queens.

:- interface.

:- import_module io.

:- pred main(io__state, io__state).
:- mode main(di, uo) is cc_multi.

:- implementation.

:- import_module list, int.

main -->
    ( \{ data(Data), queen(Data, Out) \} ->
        print_list(Out)
    ;
        io__write_string("No solution\verb+\n+")
    ).

:- pred data(list(int)).
:- mode data(out) is det.

:- pred queen(list(int), list(int)).
:- mode queen(in, out) is nondet.

:- pred qperm(list(T), list(T)).
:- mode qperm(in, out) is nondet.

:- pred qdelete(T, list(T), list(T)).
:- mode qdelete(out, in, out) is nondet.

:- pred safe(list(int)).
:- mode safe(in) is semidet.

:- pred nodiag(int, int, list(int)).
:- mode nodiag(in, in, in) is semidet.

data([1,2,3,4,5]).

queen(Data, Out) :-
    qperm(Data, Out),
    safe(Out).

qperm([], []).
qperm([X|Y], K) :-
    qdelete(U, [X|Y], Z),
    K = [U|V],
    qperm(Z, V).
\end{alltt}

\textwidth 12.7cm
\end{footnotesize}
\end{minipage}
\begin{minipage}[t]{.5\linewidth}
\textwidth 16cm
\begin{footnotesize}

\begin{alltt}
qdelete(A, [A|L], L).
qdelete(X, [A|Z], [A|R]) :-
    qdelete(X, Z, R).

safe([]).
safe([N|L]) :-
    nodiag(N, 1, L),
    safe(L).

nodiag(_, _, []).
nodiag(B, D, [N|L]) :-
    NmB is N - B,
    BmN is B - N,
    ( D = NmB ->
        fail
    ; D = BmN ->
        fail
    ;
        true
    ),
    D1 is D + 1,
    nodiag(B, D1, L).

:- mode print_list(list(int)::in, 
    io__state::di, 
    io__state::uo) is det.

print_list(Xs) -->
    ( \{ Xs = [] \} ->
        io__write_string("[]\verb+\n+")
    ;
        io__write_string("["),
        print_list_2(Xs),
        io__write_string("]\verb+\n+")
    ).

:- mode print_list_2(list(int)::in, 
    io__state::di, 
    io__state::uo) is det.

print_list_2([]) --> [].
print_list_2([X|Xs]) --> 
    io__write_int(X),
    ( \{ Xs = [] \} ->
        []
    ;
        io__write_string(", "),
        print_list_2(Xs)
 ).
\end{alltt}

\textwidth 12.7cm
\end{footnotesize}
\end{minipage}
%\caption{The N-queens program in Mercury}
\label{queens}
\vspace{0.3cm}
%\hrule
%\end{figure}

\newpage
\section*{Appendix 2 - The Mercury trace}
\label{event-section}

There are three kinds of attributes: attributes containing information relative
to the control-flow (numbered from 1 to 6 in the following), to the data-flow
(7 and 8) as well as information relative to the source code (9 and 10). The
different \emph{attributes} provided by the Mercury tracer are listed below.

\begin{enumerate} 

\item \emph{Event number} \myemph{(chrono)}. It is the rank of the event in
the trace.
\item \emph{Goal invocation number} \myemph{(call)}. 
\item \emph{Execution depth} \myemph{(depth)}. 
\item \emph{Event type or port} \myemph{(port)}.  There are the 4 traditional
  ports \myemph{call}, \myemph{exit}, \myemph{fail} and \myemph{redo}
  introduced by Byrd \cite{byrd80} for Prolog. Mercury also generates
  \emph{internal} events describing what occurs inside a call: an event of type
  \myemph{disj} is generated each time the execution enters a branch of a
  disjunction, of type \myemph{switch}\footnote {A \emph{switch} is a
    disjunction in which each branch unifies a ground variable with a different
    function symbol. In that case, at most one disjunction will provide a
    solution} if this disjunction is a switch, of type \myemph{then} if it is
  the ``then'' branch of a if-then-else and of type \myemph{else} if it is the
  ``else'' branch.

\item \emph{Determinism} \myemph{(deter)}. 
It characterizes the number of potential solutions for each procedure. The
different determinism markers are described in section \ref{mercury-pres}.

\item \emph{Procedure} \myemph{(proc)}. It is characterized by:  
a flag indicating if the current procedure is a function or a predicate  
\myemph{(proc\_type)}, \emph{a module name} \myemph{(module)}, 
\emph{a procedure name} \myemph{(proc\_name)}, \emph{an arity} \myemph{(arity)} 
and a \emph{mode number} \footnote{The mode number encodes the mode of a
predicate or a function: when a predicate has one mode, this number is 0. If
not, this number corresponds to the rank of appearance in the code of the mode
declaration; 1 for the first, 2 for the second, etc.}
\myemph{(mode\_num)}.

%   {Informations relatives aux données:} 
\item  \emph{List of live arguments} \myemph{(arg)}. A variable is said to be 
\emph{live} at a given point of the execution if its instantiation is still 
available.

\item  \emph{List of local live variables} \myemph{(local\_var)}. It is the  
live variables that are not arguments of current procedure.

\item \emph{Goal Path} 
\myemph{(goal\_path)}. It is a list indicating in which branch of the 
code the current event takes place. The branches \emph{then} and \emph{else} of
a \emph{if-then-else } are represented by \myemph{t} and
\myemph{e} respectively; the \emph{conjunctions}, \emph{disjunctions} and
the \emph{switches} are represented by \myemph{ci}, \myemph{di} and
\myemph{si} respectively, where \myemph{i} is the number of the conjunction, 
disjunction, or the switch. For example, an event whose path is
\myemph{[c3;e;d1]} corresponds to an event which occurs in the first
branch of a disjunction, which is itself part of an else branch  of an
if-then-else, which is in the third conjunction of the current procedure.
For efficiency reasons, this attribute is only available at internal events.

\item \emph{Ancestor stack} (\myemph{ancestors}).

\end{enumerate}

A more detailed description of the contents of the Mercury trace is made in the
Mercury language reference manual \cite{mercurylang99} and in the user's manual
of Morphine \cite{morphine-ref}.

\end{document}